\documentclass[11pt, letter]{article}
\usepackage{jheppub}
\usepackage[english]{babel}
\usepackage{blindtext}\blindmathtrue
\usepackage[dvipsnames]{xcolor}
\usepackage{xparse}
\usepackage{amsmath,epsf,amssymb,mathtools,latexsym,amsthm,setspace,array,pifont,hyperref,amsfonts,dsfont,cancel,braket,parskip}
\usepackage{slashed}
\usepackage{bbm}
\usepackage{graphicx}
\usepackage{tikz}
\usepackage{tikz-feynman}
\tikzfeynmanset{compat=1.0.0}

\def\XXint#1#2#3{{\setbox0=\hbox{$#1{#2#3}{\int}$}
     \vcenter{\hbox{$#2#3$}}\kern-.5\wd0}}

\usepackage[title]{appendix}

\usepackage[force]{feynmp-auto}
\usepackage{float}
\def\be{\begin{equation}}
	\def\ee{\end{equation}}
\def\beA{\begin{align}}
\def\eeA{\end{align}}
	
\newcommand{\del}{\partial}	
\newcommand{\nn}{\nonumber}

\newcommand{\M}{\mathcal{M}}

\newcommand{\Tr}{\text{Tr}}

\def\prd{\ref@{Phys.~Rev.~D}}        

\title{Gauge invariance and generalised  $\eta$ regularisation}

\author{Antonio Padilla}
\author{and Robert G. C. Smith}

\affiliation{School of Physics and Astronomy, University of Nottingham, University Park, Nottingham NG7 2RD, United Kingdom}
\affiliation{The Nottingham Centre of Gravity, University of Nottingham, Nottingham NG7 2RD, UK}

\emailAdd{antonio.padilla@nottingham.ac.uk}
\emailAdd{robert.smith@nottingham.ac.uk}
\abstract{We generalise the $\eta$ regularisation scheme in order to develop a framework for systematically studying regularisation of loops in quantum field theory.  This allows us to``solve" a set of gauge consistency conditions for families of gauge invariant regularisation schemes. We recover several known examples such as dimensional  and denominator regularisations, as well as some more general solutions.  We also  study anomalies in chiral theories in order to carefully describe how our formalism should be properly implemented. }

\begin{document}
\maketitle
\section{Introduction}
In perturbative quantum field theory (QFT), renormalisation is fundamental to our ability to compute measurable quantities. The essence of renormalisation is to utilise a regularisation prescription that can consistently isolate and remove all infinities. Without a consistent and robust regularisation method, these infinities would otherwise prohibit any systematic predictions from loop amplitudes.

There are a number of common regularisation schemes, all of which have have their advantages and disadvantages \cite{Bertlmann:1996xk,Schwartz14}. When evaluating the strengths and weaknesses of a particular regularisation method, an important criterion is the preservation of gauge symmetry.    All of the fundamental forces of nature - i.e., gravitational, electromagnetic, weak and strong forces - are governed by basic symmetry principles. For example, the electromagnetic, weak and strong forces are characterised by the gauge symmetry $U(1) \times SU(2) \times SU(3)$, while the gravitational force is governed by diffeomorphism invariance. In fact, any Lorentz invariant theory of massless fields of spin one or spin two must possess a gauge symmetry \cite{Weinberg95a}.  

The calculation of perturbative quantum corrections can also give rise to quantum anomalies \cite{Bertlmann:1996xk, Fujikawa:2004cx, bilal2008lectures}. A quantum anomaly is the failure to preserve a classical symmetry in the process of quantisation and regularisation.  This is not necessarily a problem if the classical symmetry is a global one. On the other hand, if the classical symmetry is local, a gauge anomaly generates a mismatch between the numbers of the degrees of freedom in the classical and quantum corrected theories. 

Chiral anomalies are the prototypical example often used to probe the effects of regularisation,  first discovered in the context of spinor electrodynamics.  Classically, there are two conserved currents: an axial current associated with the chiral symmetry and a vector current associated with the gauge symmetry.   In 1969 Adler \cite{Adler:1969gk} and Bell and Jackiw \cite{Bell:1969ts}, showed in the computation of decay processes of a neutral pion to two photons that  only the vector current is conserved after quantum corrections are taken into account. As a result, the conservation of the axial current is found to be anomalously broken. 

Of course, not every regularisation scheme is well-suited to studying a particular process for a particular QFT. For example, Pauli-Villars regularisation runs into issues with broken gauge invariance, in contrast to dimensional regularisation \cite{tHooft:1972tcz}. For its part, dimensional regularisation is hard to understand intuitively and is tricky to implement in chiral theories due the presence of the dimensionally dependent definition of the  $\gamma_5$ matrix.

In this paper, we study the problem of gauge invariance in a scheme independent way and in a fixed number of dimensions. The tools for doing this are inspired by the $\eta$ regularisation scheme developed in \cite{PadillaSmith24}.  $\eta$ regularisation is a generalisation to QFT of the method of smoothed asymptotics studied in number theory\cite{Tao11}. Just as smoothed asymptotics offers an intuitive and technically simple approach to the regularisation of divergent series, so $\eta$ regularisation does the same for divergent loop integrals.  This technical simplicity allowed us to reveal an unexpected relationship between the regularisation of divergent series in number theory and gauge invariance in QFT.

We shall generalise $\eta$ regularisation just enough to capture most known regularisation schemes, whilst retaining enough simplicity to ``solve" a set of gauge consistency conditions. These ``solutions" correspond to regularisation schemes that preserve gauge invariance, like dimensional regularisation.  We also study anomalies in chiral theories within this generalised framework. This allows us to demonstrate how $\eta$ regularisation ought to be properly implemented in order to recover the correct results.  

This is a first step towards a generalised theory of consistent regulators which may offer a tantelising glimpse of something more fundamental. Indeed, if string theory or some other fundamental theory is UV finite, it could point towards a preferred regularisation scheme in the field theory limit.  It is this interplay between regularisation, consistency and UV finiteness that we wish to explore in a series of papers.

The rest of this paper is organised as follows: in the next section we give an introduction to $\eta$ regularisation, generalising the underlying ideas presented in \cite{PadillaSmith24}.  We emphasize  the role of momentum routing and how this connects to gauge invariance.  In section \ref{sec:gauge},  we explain how one can go about solving gauge consistency conditions for the regulator. We are able to recover a host of gauge invariant regularisation schemes, some of which were already known, some of which are new.   In section \ref{triangle} we study anomalies in chiral theories before ending with a discussion in section \ref{discussion}. 

\section{Generalised $\eta$ regularisation}

A systematic study of regularisation in QFT benefits from a canonical form for the diverging integrals that appear in a generic theory. At one loop and for QFTs defined on Minkowski space, this canonical form is provided by the one-fold irreducible loop integrals (ILIs), as advocated by Wu \cite{Wu03,Wu04,Wu14} but also discussed in earlier works \cite{Battistel:1998sz}. In four spacetime dimensions, the ILIs of integer spin $s$ are given by

\begin{equation}
I^{\mu_1 \cdots \mu_s}_{-2 \alpha} (\M^2) = 
		\int \frac{d^4 k}{(2 \pi)^4}  \frac{k^{\mu_1}\cdots k^{\mu_s}}{(k^2 +\M^2-i \epsilon)^{2+\frac{s}{2}+\alpha}} 
\end{equation}

The subscript $(-2\alpha)$ labels the power counting dimension of energy-momentum with  $\alpha = -1$ and $\alpha = 0$ corresponding  to quadratic and logarithmically divergent integrals. The mass term  $\M^2 =\M^2(m^2_1, p_1^2, \dots) $ is a function of Feynman parameters, external momenta $p_i$, and corresponding mass scales $m_i$. We define  $k^2=g_{\mu\nu} k^\mu k^\nu$ where the metric $g_{\mu\nu}$ is written with mostly positive signature. To generalise this canonical form to $n$ loops, one can introduce the $n$-fold ILIs for which there are no longer the overlapping factors $(k_i - k_j + p)$ in the denominator of the integrand and no factors of the scalar momentum  $k^2$ in the numerator \cite{Wu14}.

Following a similar strategy to \cite{PadillaSmith24}, we implement a generalised form of $\eta$ regularisation by Wick rotating to Euclidean signature $k_0 \to i k_4$  and inserting a regularisation factor of $\eta\left(\frac{k}{\mu}, \epsilon\right)$ into the integrand of the ILI, where $k$ is now the norm of the Euclidean four-momentum, $\mu$ is a fixed mass scale, and $\epsilon$ is a dimensionless parameter that controls the rate at which the regulator damps the corresponding integrals. This  yields $I_{-2 \alpha }^{\cdots }\to iJ_{-2\alpha}^{\cdots}[\eta]$,  where the $\eta$ factor need not coincide for different ILIs of different spin. We will see why this must be the case in a moment.  In any event, the regularised ILIs are given by 
\begin{equation}
J^{\mu_1 \cdots \mu_s}_{-2 \alpha} [\eta] (\M^2) = 
		\int \frac{d^4 k}{(2 \pi)^4}  \frac{k^{\mu_1}\cdots k^{\mu_s}}{(k^2 +\M^2)^{2+\frac{s}{2}+\alpha}} \eta\left(\frac{k}{\mu}, \epsilon\right)
\end{equation}

where the integration is over four-dimensional Euclidean space. Note that the rotational symmetry ensures that the ILIs of odd spin vanish identically. 

As in  \cite{PadillaSmith24}, the role of the regulator function $\eta$ is to suppress the ultraviolet divergences in the integrals.  In  \cite{PadillaSmith24} we took $\eta =\eta(k/\Lambda)$ where $\Lambda$ is some regularisation scale and $\eta(x)$ was assumed to be a smooth function defined on the positive real line.  For this to correspond to a good regulator, we further assumed that  $\eta(x) \to 1$ as $x\to 0^+$ and that $\eta(x)$ was a Schwartz function, vanishing faster than any negative power of $x$ as $x \to \infty$.   In this paper we explore a more general form for our regulator functions, namely $\eta\left(\frac{k}{\mu}, \epsilon\right)$ as described above. For $\eta(y, \epsilon)$ to be a good regulator function, we now  require it to be smooth in $y$ and to satisfy the following: 
\begin{itemize}
\item in a neighbourhood of $\epsilon=0$, we have $\eta(y, \epsilon) \approx 1+\epsilon^\delta f(y)$ for some $\delta>0$ and some smooth function $f(y)$; 
\item there exists $a, b, c >0$, such that for  $\epsilon\geq a$ , we have $|\eta(y, \epsilon)|< c y^{-2b}$ as $y \to \infty$. 
\end{itemize}
The first condition states that the regulator function tends smoothly to one as we remove the control parameter, $\epsilon$.  The second condition states that for $\epsilon$ greater than some critical value $a>0$, the regulator dies off at infinity faster than some negative power.  Clearly this is a much weaker set of conditions than those imposed in \cite{PadillaSmith24}.  For example, we can now capture the form of the regulator expected from dimensional regularisation, for which $\eta(y, \epsilon)=y^{-2\epsilon}$.  If we were to identify a regularisation scale, $\Lambda=\mu/\epsilon$, the class of regulator functions studied in \cite{PadillaSmith24} correspond to the case where $\eta(y, \epsilon)=\eta(y \epsilon)$. 

Whenever we compute a correlation function up to one loop in a given QFT,  we have to do some work to write it in terms of ILIs.  Many of these manipulations are only valid if the integrals are regulated. These regularisations should be implicitly understood such that we arrive at the desired form of regularised ILIs, consistent with our generalised $\eta$ framework. In this sense, our approach is similar in spirit to the implicit regularisation methods introduced in \cite{Battistel:1998sz}.  Of course, some manipulations involve the use of Feynman parametrisation \cite{Weinberg95a}. In the appendix, we explicitly show how the regulator function changes form as we undo a Feynman parametrisation.   

The last step of Feynman parametrisation is always to shift the momentum integration.  This might be a worry, since it is not at all  obvious that momentum routing is allowed under $\eta$ regularisation. In  \cite{PadillaSmith24}, we allowed the regulators to differ for integrals of different spin. This was argued for on the grounds of generality, although we will now show that it was crucial to allow for momentum routing.

Consider the following regularised ILIs and impose momentum routing, with $\rho\left(\frac{k}{\mu}, \epsilon\right)$ a general smooth regulator,
\begin{eqnarray}
J^{\mu_1 \cdots \mu_s}_{-2\alpha}[\rho_{-2\alpha}] (\M^2) &&=\int \frac{d^4 k}{(2 \pi)^4}  \frac{k^{\mu_1}\cdots k^{\mu_s}}{(k^2 +\M^2)^{2+\frac{s}{2}+\alpha}} \rho_{-2\alpha}\left(\frac{k}{\mu}, \epsilon\right)   \\
&&=\int \frac{d^4 k}{(2 \pi)^4}  \frac{(k+q)^{\mu_1}\cdots (k+q)^{\mu_s}}{((k+q)^2 +\M^2)^{2+\frac{s}{2}+\alpha}} \rho_{-2\alpha}\left(\frac{|k+q|}{\mu}, \epsilon\right)  \nn \\
&&=\int \frac{d^4 k}{(2 \pi)^4} e^{q^\nu\frac{\partial}{\partial k^\nu}} \left[ \frac{k^{\mu_1}\cdots k^{\mu_s}}{(k^2 +\M^2)^{2+\frac{s}{2}+\alpha}}  \rho_{-2\alpha}\left(\frac{k}{\mu}, \epsilon\right)\right] \nn
\end{eqnarray}
For this to be true for any value of $q$, we must have the following constraints for any  integer value of $r, s \geq 0$
\begin{equation}
 \int \frac{d^4 k}{(2 \pi)^4} \frac{\partial^s}{\partial k^{\nu_1} \cdots \partial k^{\nu_r} }	       \left[ \frac{k^{\mu_1}\cdots k^{\mu_s}}{(k^2 +\M^2)^{2+\frac{s}{2}+\alpha}}  \rho_{-2\alpha}\left(\frac{k}{\mu}, \epsilon\right)\right] =0
\end{equation}
Thanks to the rotational symmetry of the scalar integrand, this is trivially true for odd values of $r+s$. However, for even values, we obtain  non-trivial relations between ILIs of different spin.  For example, for $r=s=1$ and taking $\rho=\eta_{-2\alpha}$, we see that
\begin{equation}
 \int \frac{d^4 k}{(2 \pi)^4} \frac{\partial}{\partial k^{\mu}  }        \left[  \frac{k_{\nu}}{(k^2 +\M^2)^{2 + \alpha}}\eta_{-2\alpha}\left(\frac{k}{\mu}, \epsilon\right) \right]=0
\end{equation}

For $\alpha \neq -2$, this yields the following relationship between the ILIs of spin two and spin zero,
\begin{equation}\label{spin2cond}
J_{-2\alpha}^{\mu\nu}[\theta_{-2\alpha}] (\M^2)=\frac{1}{2(2+\alpha)} g^{\mu\nu}J_{-2\alpha}[\eta_{-2\alpha}] (\M^2) 
\end{equation}
 where 
 \begin{equation} \label{theta}
 \theta_{-2\alpha}\left(y, \epsilon\right)=-\frac{(y^2+y_0^2)^{3+\alpha}}{2(2+\alpha)}  \frac{\del}{y\del y} \left[\frac{\eta_{-2\alpha}\left(y, \epsilon\right)}{(y^2+y_0^2)^{2+\alpha}}\right]
 \end{equation}
Similarly,  for $r=s=2$ and again taking $\rho=\eta_{-2\alpha}$, we find 
 \begin{equation}
 \int \frac{d^4 k}{(2 \pi)^4} \frac{\partial^2}{\partial k^{\mu_1}\partial k^{\mu_2}  }        \left[  \frac{k_{\mu_3}k_{\mu_4}}{(k^2 +\M^2)^{2 + \alpha}}\eta_{-2\alpha}\left(\frac{k}{\mu}, \epsilon\right) \right]=0.
\end{equation}
 For $\alpha \neq -2, -3$,  this gives a complex relationship between the ILIs of  spin four, spin two and spin zero, 
 \begin{multline} \label{spin4condcomplex}
J_{-2\alpha}^{\mu_1 \cdots \mu_4}[\kappa_{-2\alpha}] =\frac{1}{2(3+\alpha)}\left[
J_{-2\alpha}^{\mu_1 \mu_3} [\theta_{-2\alpha}]  g^{\mu_2\mu_4}
+J_{-2\alpha}^{\mu_1 \mu_4} [\theta_{-2\alpha}]  g^{\mu_2\mu_3}
+J_{-2\alpha}^{\mu_2 \mu_3} [\theta_{-2\alpha}]  g^{\mu_1\mu_4} \right.\\
\left.
+J_{-2\alpha}^{\mu_2 \mu_4} [\theta_{-2\alpha}]  g^{\mu_1\mu_3}
+J_{-2\alpha}^{\mu_3 \mu_4} [\eta_{-2\alpha}]  g^{\mu_1\mu_2}
 \right]
-\frac{1}{4(2+\alpha)(3+\alpha)}J_{-2\alpha}[\eta_{-2\alpha}]( g^{\mu_1\mu_3}g^{\mu_2\mu_4}+ g^{\mu_1\mu_4} g^{\mu_2\mu_3})
\end{multline}
 where 
 \begin{equation} \label{kappa}
 \kappa_{-2\alpha}\left(y, \epsilon\right)=
\frac{(y^2+y_0^2)^{4+\alpha}}{4(2+\alpha)(3+\alpha)}  \left(\frac{\del}{y\del y}\right)^2 \left[\frac{\eta_{-2\alpha}\left(y, \epsilon\right)}{(y^2+y_0^2)^{2+\alpha}}\right]
 \end{equation}
 Combining \eqref{spin4condcomplex} with \eqref{spin2cond}, we can by-pass the ILIs of spin 2, and relate the ILIs of spin four and spin zero directly, 
 \begin{equation}\label{spin4cond}
J_{-2\alpha}^{\mu_1 \cdots \mu_4}[\kappa_{-2\alpha}]  (\M^2)=\frac{1}{4(2+\alpha)(3+\alpha)} S^{\mu_1 \cdots \mu_4} J_{-2\alpha}[\eta_{-2\alpha}] (\M^2) 
\end{equation}
 where $S^{\mu_1 \cdots \mu_4}=g^{\mu_1\mu_2}g^{\mu_3\mu_4}+g^{\mu_1\mu_3}g^{\mu_2\mu_4}+g^{\mu_1\mu_4}g^{\mu_2\mu_3}$ is the symmetric product of metrics. 
 
 It is not too difficult to convince oneself that if  $\eta_{-2\alpha}$  is a good regulator in the way we defined earlier, then the same is true of  both   $\theta_{-2\alpha}$ and $\kappa_{-2\alpha}$. Furthermore, we note that \eqref{spin2cond} and \eqref{spin4cond} are identical to the gauge consistency conditions derived by Wu \cite{Wu03,Wu04, Wu14} from generalised Ward identities in a generic gauge theory with  a gauge group of arbitrary rank and an arbitrary number of Dirac spinors.  This alignment is no coincidence, as discussed in detail in \cite{Battistel:1998sz}. In \cite{PadillaSmith24}, we imposed those gauge consistency conditions in $\eta$ regularisation while allowing the regulator to differ for integrals of different spin and different dimension.  We now see that this was a necessary ingredient, given the explicit form of $\theta_{-2\alpha}$ and $\kappa_{-2\alpha}$ presented in \eqref{theta} and \eqref{kappa} derived using momentum routing.
 
We end this section with a brief comment on causality.  As is well known, to preserve the causal structure, we should implement  Feynman's $i \epsilon$ prescription prior to any Wick rotation from Minkowski to Euclidean signature,  $k_0 \to i k_4$.  In a generalised $\eta$ regularisation scheme, it should be possible to Wick rotate the integrand as long as the smooth regulator functions do not introduce any extra poles that block the path of the contour rotation.  

\section{Gauge invariant $\eta$ regularisation} \label{sec:gauge}
$\eta$ regularisation provides us with a formalism to study regularisation schemes in an abstract way.  To illustrate this, we consider the question of gauge invariance. It is well known that many popular regularisation schemes such as cut-off regularisation or Pauli-Villars can run into problems with broken gauge invariance.  In this section, we will show how one can use  the formalism of $\eta$ regularisation to ``solve" for regularisation schemes that preserve the underlying gauge symmetries. 

Consider a generic gauge theory where the gauge group has dimension $d_G$ and where there are $N_f$ Dirac spinors $\psi_n$ ($n=1, \ldots, N_f$) interacting with the Yang Mills field $A_\mu^a$ ($a=1, \ldots , d_G$).  This is described by a Lagrangian, 
\be \label{lagr}
\mathcal{L}=\bar \psi_n (i\gamma^\mu D_\mu-m)\psi_n -\frac14 F_{\mu\nu}^a F^{\mu\nu}_a,
\ee
where
\be
F^a_{\mu\nu}=\partial_{\mu} A_\nu^a-\partial_{\nu} A_\mu^a - gf_{abc} A^b_\mu A^c_\nu, \qquad D_\mu \psi_n=(\partial_\mu +ig T^a A^a_\mu )\psi_n,
\ee
and  $T^a$ are the generators of the gauge group whose commutator $[T^a, T^b]=i f^{abc}T^c$ defines the structure constants $f^{abc}$.  Following \cite{Wu03,Wu04, Wu05} and discussed further in \cite{PadillaSmith24},  one can compute correlation functions at one loop and impose a set of generalised Ward identities.  For example, the vacuum polarisation of the gauge field takes the form,
\be
\Pi^{ab}_{\mu\nu}(p)=\Pi^{(g) ab}_{\mu\nu}(p)+\Pi^{(f) ab}_{\mu\nu}(p),
\ee
where $p^\mu$ is the external momentum, $\Pi^{(g) ab}_{\mu\nu}(p)$ are the pure Yang Mills contributions coming from  gauge field loops and ghost loops, and  $\Pi^{(f) ab}_{\mu\nu}(p)$ are the contributions from fermion loops, coming from the interaction of the fermions with the gauge field.  As is well known,  gauge invariance is given in terms of the Ward identities $p^\mu\Pi^{ab}_{\mu\nu}=\Pi^{ab}_{\mu\nu}p^\nu=0$.  However,  since our interest lies in the regularisation scheme rather than the details of any individual theory, we require these Ward identities to hold for {\it any} gauge theory and with {\it any} number of fermions. This suggests that the Ward identities should hold separately for the gauge field and fermionic contributions
\be \label{condition}
 p^\mu\Pi^{{(g)} ab}_{\mu\nu}=\Pi^{(g)ab}_{\mu\nu}p^\nu=0, \qquad  p^\mu\Pi^{{(f)} ab}_{\mu\nu}=\Pi^{(f)ab}_{\mu\nu}p^\nu=0.
\ee
One can apply a similar logic to the Ward identities derived from other correlation functions.  In order to satisfy the resulting generalised Ward identities, we arrive at consistency conditions on the ILIs for $\alpha=-1, 0, 1$, which are those derived from momentum routing, in \eqref{spin2cond} and \eqref{spin4cond}. Note that for $\alpha>0$, the integrals are convergent and no regulator is required. In this case, the consistency conditions are satisfied trivially as,  of course, they should be.  For this, our remaining focus is on the case where $\alpha=0, -1$, corresponding to logarithmic and quadratically divergent ILIs, respectively.

For a suitably chosen $\eta_{-2\alpha}$, we may be tempted to regard \eqref{theta} and \eqref{kappa} as ``solutions" of the gauge consistency conditions. However, these solutions are not particularly useful or illuminating. The reason is that one is always free to shift each of the regulators as follows, without having any impact on the consistency conditions \eqref{spin2cond} and \eqref{spin4cond}, 
\begin{eqnarray}
\eta_{-2\alpha}\left(y, \epsilon\right) \ & \to & \eta_{-2\alpha}\left(y, \epsilon\right) + Z^{(0)}_{-2\alpha}\left(y, \epsilon\right) \label{shift1}\\
\theta_{-2\alpha}\left(y, \epsilon\right) \ & \to & \theta_{-2\alpha}\left(y, \epsilon\right) + Z^{(2)}_{-2\alpha}\left(y, \epsilon\right) \label{shift2} \\
\kappa_{-2\alpha}\left(y, \epsilon\right) \ & \to & \kappa_{-2\alpha}\left(y, \epsilon\right) + Z^{(4)}_{-2\alpha}\left(y, \epsilon\right) \label{shift3}
\end{eqnarray}
where the $Z^{(n)}$ for $n=0, 2, 4$ are smooth functions defined on $y\geq 0$. These functions should vanish in the asymptotic limit,  $Z^{(n)}(y, \epsilon) \to 0$ as $\epsilon \to 0$. For sufficiently large $\epsilon>0$, they should decay at large momenta,  vanishing faster than some inverse polynomial as $y \to \infty$.  These conditions ensure that the shifts given above preserve the properties of a good regulator defined in the previous section. However, in order to also protect the value of the ILIs,  and by association the gauge consistency conditions, we further require
\begin{equation}
J_{-2\alpha}[Z^{(0)}_{-2\alpha}] (\M^2) =0, \qquad J_{-2\alpha}^{\mu\nu}[Z^{(2)}_{-2\alpha}] (\M^2) =0, \qquad J_{-2\alpha}^{\mu_1 \ldots \mu_4\nu}[Z^{(4)}_{-2\alpha}] (\M^2) =0.
\end{equation}
Note that  a shift in, say,  $\eta_{-2 \alpha}$ of the form \eqref{shift1} will generate a shift of the form \eqref{shift2} and \eqref{shift3} in the $\theta_{-2\alpha}$ and $\kappa_{-2\alpha}$ under the definition \eqref{theta} and \eqref{kappa}.

All of this suggests that the consistency equations \eqref{spin2cond} and \eqref{spin4cond} are equations for the regulators that may be solved directly, allowing for solutions that do not necessarily take the specific form given by \eqref{theta} and \eqref{kappa}.  This is the approach we will take to finding consistent regulators in a familiar form.

To this end, we use the rotational symmetry to write the regulated ILIs for spin two and spin four as scalar integrals, 
\begin{eqnarray}
J_{-2\alpha}^{\mu\nu}[\theta_{-2\alpha}] (\M^2) &=& \frac{1}{4} g^{\mu\nu} \int \frac{d^4 k}{(2 \pi)^4}  \frac{k^2}{(k^2 +\M^2)^{3+\alpha}} \theta_{-2\alpha} \left(\frac{k}{\mu}, \epsilon\right)\\
J_{-2\alpha}^{\mu_1 \cdots \mu_4}[\kappa_{-2\alpha}]  (\M^2) &=&\frac{1}{4!} S^{\mu_1 \cdots \mu_4} \int \frac{d^4 k}{(2 \pi)^4}  \frac{k^4}{(k^2 +\M^2)^{4+\alpha}} \kappa_{-2\alpha} \left(\frac{k}{\mu}, \epsilon\right)\
\end{eqnarray}
Plugging these expressions into the conditions \eqref{spin2cond} and \eqref{spin4cond} and rearranging a little, we obtain the following integral constraints
\begin{eqnarray}
\int_0^\infty dy \   \frac{y^3} {(y^2+y_0^2)^{2+\alpha}} \left[  \frac{y^2}{y^2+y_0^2} \theta_{-2\alpha}-\frac{2}{2+\alpha}\eta_{-2\alpha}  \right] &=& 0  \label{int1}\\
\int_0^\infty dy \   \frac{y^5} {(y^2+y_0^2)^{3+\alpha}} \left[  \frac{y^2}{y^2+y_0^2} \kappa_{-2\alpha}-\frac{3}{3+\alpha}\theta_{-2\alpha}  \right] &=& 0 \label{int2}
\end{eqnarray}
It will be useful to construct solutions to these constraints in the form
\begin{eqnarray}
\frac{y^2}{y^2+y_0^2} \theta_{-2\alpha}-\frac{2}{2+\alpha}\eta_{-2\alpha} &=& \frac{(y^2+y_0^2)^{2+\alpha}}{y^3} \frac{\del }{\del y} \psi_{-2\alpha}(y, \epsilon)  \label{psi}\\
\frac{y^2}{(y^2+y_0^2)} \kappa_{-2\alpha}-\frac{3}{(3+\alpha)}\theta_{-2\alpha} &=& \frac{(y^2+y_0^2)^{3+\alpha}}{y^5} \frac{\del }{\del y} \phi_{-2\alpha}(y, \epsilon) \label{phi}
\end{eqnarray}
where $\psi_{-2\alpha}(y, \epsilon)$ and $\phi_{-2\alpha}(y, \epsilon)$ are smooth functions defined on $y\geq 0$.  For  sufficiently large  $\epsilon>0$, these functions should vanish at $y=0$ and as  $y \to \infty$. Furthermore, in the limit where $\epsilon \to 0$, recall that  all regulators tend to unity, from which we infer that
\begin{equation}
\psi_{-2\alpha}(y, 0)=-\frac{1}{2(2+\alpha)} \frac{y^4}{(y^2+y_0^2)^{2+\alpha}} ,
\qquad
\phi_{-2\alpha}(y,0)=-\frac{1}{2(3+\alpha)} \frac{y^6}{(y^2+y_0^2)^{3+\alpha}}  \label{eps=0}
\end{equation}
where we have ignored  the possibility of an irrelevant additive constant in each case.  
\subsection{Explicit ``solutions"}
We are now ready to state explicit solutions to the consistency conditions \eqref{spin2cond} and \eqref{spin4cond}  using the solution generating method we have just derived. 

\paragraph{Tao inspired regularisation schemes:}

We begin with the regularisation schemes explored in \cite{PadillaSmith24}, for which the regulators have the form $\eta=\eta(\epsilon y)$.  Introducing $x=\epsilon y$, the integral constraints \eqref{int1} and \eqref{int2} now read
\begin{eqnarray}
\epsilon^{2\alpha}\int_0^\infty dx \   \frac{x^3} {(x^2+\epsilon^2 y_0^2)^{2+\alpha}} \left[  \frac{x^2}{x^2+\epsilon^2 y_0^2} \theta_{-2\alpha}-\frac{2}{2+\alpha}\eta_{-2\alpha}  \right] &=& 0  \label{int1x}\\
\epsilon^{2\alpha} \int_0^\infty dx \   \frac{x^5} {(x^2+ \epsilon^2 y_0^2)^{3+\alpha}} \left[  \frac{x^2}{x^2+ \epsilon^2 y_0^2} \kappa_{-2\alpha}-\frac{3}{3+\alpha}\theta_{-2\alpha}  \right] &=& 0 \label{int2x}
\end{eqnarray}
If we carry out an expansion in  $\epsilon \ll 1$, we obtain 
\begin{eqnarray}
&& \epsilon^{2\alpha}\int_0^\infty dx \   x^{-1-2\alpha} \left[ \theta_{-2\alpha}-\frac{2}{2+\alpha}\eta_{-2\alpha}  \right]  \nn \\
&& \qquad +\epsilon^{2(\alpha+1)}y_0^2 \int_0^\infty dx \   x^{-3-2\alpha} \left[ -(3+\alpha) \theta_{-2\alpha}+2\eta_{-2\alpha}  \right] +\mathcal{O}\left(\epsilon^{2(\alpha+2)}\right) = 0  \label{int1xexp}\\
&&\epsilon^{2\alpha} \int_0^\infty dx \   x^{-1-2\alpha} \left[  \kappa_{-2\alpha}-\frac{3}{3+\alpha}\theta_{-2\alpha}  \right] \nn \\
&& \qquad +\epsilon^{2(\alpha+1)}y_0^2 \int_0^\infty dx \   x^{-3-2\alpha} \left[ -(4+\alpha) \kappa_{-2\alpha}+3\theta_{-2\alpha}  \right] +\mathcal{O}\left(\epsilon^{2(\alpha+2)}\right) = 0 \label{int2xexp}
\end{eqnarray}
where we have explicitly shown the terms which are non-vanishing as $\epsilon  \to 0$, at least for $\alpha=0, -1$.  For quadratically divergent ILIs with $\alpha=-1$, requiring the divergent parts to vanish order by order in $\epsilon$ gives the following constraints,
\begin{eqnarray}
&& \int_0^\infty dx \   x \left[ \theta_{2}-2\eta_{2}  \right]  = 0,  \qquad  \int_0^\infty dx \   x \left[ 2 \kappa_{2}-3\theta_{2}  \right] = 0  \label{p1con1} \\
 &&\int_0^\infty dx \  \frac{ \theta_{2}-\eta_{2}}{x}   =0,  \qquad 
\int_0^\infty dx \   \frac{ \kappa_{2}-\theta_{2} }{x}  = 0  \label{p1con2}
\end{eqnarray}
Similarly, for logarithmically divergent ILIs with $\alpha=0$, we obtain  
\begin{equation}
\int_0^\infty dx \  \frac{ \theta_{0}-\eta_{0}}{x}   =0,  \qquad 
\int_0^\infty dx \   \frac{ \kappa_{0}-\theta_{0} }{x}  = 0  \label{p1con3}
\end{equation}
The expressions \eqref{p1con1} to \eqref{p1con3} are exactly equivalent to the ``enhanced regulator" conditions derived in \cite{PadillaSmith24}.  Recall that the vanishing Mellin transform given by \eqref{p1con1} connects gauge invariance in this scheme to the suppression of divergences in number theory.

 \paragraph{Dimensional regularisation:}
 We now turn our attention to a more familiar regularisation scheme and demonstrate how it can emerge in our formalism.   In dimensional regularisation, the ILIs take the following form
 \begin{eqnarray}
J_{-2\alpha} |_\text{dim reg}[\eta_{-2\alpha}] &=& \mu^{2\epsilon} \int \frac{d^{4-2\epsilon}k}{(2\pi)^{4-2\epsilon}} \frac{1}{(k^2 + \M^2)^{2+\alpha}}\nn\\
&=& \int \frac{d^4 k}{(2\pi)^{4} }f(\epsilon)\left(\frac{k}{\mu}\right)^{-2\epsilon} \frac{1}{(k^2 + \M^2)^{2+\alpha}} \\
J_{-2\alpha}^{\mu \nu} |_\text{dim reg}[\theta_{-2\alpha}]  &=& \mu^{2\epsilon} \int \frac{d^{4-2\epsilon}k}{(2\pi)^{4}} \frac{k^{\mu}k^{\nu}}{(k^2 + \M^2)^{3+\alpha}}    \qquad \nn  \\
&=& \frac{\mu^{2\epsilon}}{4-2\epsilon} g_{\mu \nu} \int \frac{d^{4-2\epsilon}k}{(2\pi)^4} \frac{k^2}{(k^2 + \M^2)^{3+\alpha}} \nn \\
 &=&\frac14  g^{\mu \nu} \int \frac{d^{4}k}{(2\pi)^4} \frac{f(\epsilon)}{\left(1-\frac{\epsilon}{2}\right)} \left(\frac{k}{\mu}\right)^{-2\epsilon}\frac{k^2}{(k^2 + \M^2)^{3+\alpha}} \\
 J_{-2\alpha}^{\mu_1 \cdots \mu_4}|_\text{dim reg}[\kappa_{-2\alpha}]  (\M^2) 
 &=& \mu^{2\epsilon}   \int \frac{d^{4k-2\epsilon}}{(2 \pi)^{4-2\epsilon}}  \frac{k^{\mu_1} \cdots k^{\mu_4}}{(k^2 +\M^2)^{4+\alpha}} \nn \\
&=&  \frac{1}{(4-2\epsilon)(6-2 \epsilon)} S^{\mu_1 \cdots \mu_4} \int \frac{d^{4-2\epsilon} k}{(2 \pi)^{4-2\epsilon}}  \frac{k^4}{(k^2 +\M^2)^{4+\alpha}} \nn \\
&=&  \frac{1}{4!}S^{\mu_1 \cdots \mu_4} \int \frac{d^{4} k}{(2 \pi)^{4}}  \frac{f(\epsilon)}{\left(1-\frac{\epsilon}{2}\right)\left(1-\frac{\epsilon}{3}\right)}\left(\frac{k}{\mu}\right)^{-2\epsilon}  \frac{k^4}{(k^2 +\M^2)^{4+\alpha}} \qquad \quad 
\end{eqnarray}
where 
\begin{equation}
f(\epsilon)=\frac{(2\pi)^\epsilon \Omega_{3-2\epsilon}}{\Omega_3}
\end{equation}
 and $\Omega_n=2 \pi^\frac{n}{2}/\Gamma\left(\frac{n}{2}\right)$ is the volume of a unit $n$-sphere.  In other words, 
 \begin{equation}
\eta_{-2\alpha}=f(\epsilon) y^{-2\epsilon} , \qquad \theta_{-2\alpha}=\frac{f(\epsilon) y^{-2\epsilon} }{1-\frac{\epsilon}{2}}, \qquad \kappa_{-2\alpha}=\frac{f(\epsilon) y^{-2\epsilon}}{\left(1-\frac{\epsilon}{2}\right)\left(1-\frac{\epsilon}{3}\right)} \label{dimregsol}
\end{equation}
 To see how this solution might have been obtained from the generating technique described above, we consider the smooth seed functions introduced in \eqref{psi} and \eqref{phi}, and make the following simple choice
 \be\label{Etadimreg1}
\psi_{-2\alpha}(y, \epsilon)=-\frac{A(\epsilon)}{2(2+\alpha)} \frac{y^{4-2 \epsilon}}{(y^2+y_0^2)^{2+\alpha}} ,
\qquad
\phi_{-2\alpha}(y,\epsilon)=-\frac{B(\epsilon)}{2(3+\alpha)} \frac{y^{6-2 \epsilon} }{(y^2+y_0^2)^{3+\alpha}}, 
\end{equation}
where the coefficients $A, B$ depend on the regularisation parameter $\epsilon$ and are assumed to behave smoothly at the origin such that $A(0) = B(0)=1$.  It is easy to see that this choice satisfies all the conditions we impose on the seed functions above.  It might also have been easily guessed as a deformation of the  known solution at $\epsilon=0$.  Plugging \eqref{Etadimreg1} into \eqref{psi} and \eqref{phi}, we find that
\begin{eqnarray}
\frac{y^2}{y^2+y_0^2} \theta_{-2\alpha}-\frac{2}{2+\alpha}\eta_{-2\alpha} &=& \frac{y^2}{y^2+y_0^2}A(\epsilon) y^{-2\epsilon} -\frac{2}{2+\alpha}A(\epsilon)\left(1-\frac{\epsilon}{2}\right) y^{-2\epsilon} \label{psidimreg}\\
\frac{y^2}{(y^2+y_0^2)} \kappa_{-2\alpha}-\frac{3}{(3+\alpha)}\theta_{-2\alpha} &=&\frac{y^2}{(y^2+y_0^2)}B(\epsilon) y^{-2\epsilon}-\frac{3}{(3+\alpha)}B(\epsilon)\left(1-\frac{\epsilon}{3}\right) y^{-2\epsilon} \label{phidimreg}
\end{eqnarray}
It is easy to read off the solution \eqref{dimregsol} provided we identify $A(\epsilon)=f(\epsilon)/\left(1-\frac{\epsilon}{2}\right)$ and $B(\epsilon)=f(\epsilon)/\left(1-\frac{\epsilon}{2}\right)\left(1-\frac{\epsilon}{3}\right)$.  Notice that the solution generating technique correctly identifies the form of the regulators in  \eqref{dimregsol}  and their ratios. 

\paragraph{Denominator regularisation:}
Another gauge invariant regularisation scheme is  denominator regularisation \cite{Horowitz:2022rpp,Horowitz22,Bansal:2022juh,Horowitz:2023xyd}. In this case the ILIs take the form
 \begin{eqnarray}
J_{-2\alpha} |_\text{den reg}[\eta_{-2\alpha}] &=& \mu^{2\epsilon} f_{(2+\alpha, -2\alpha)}(\epsilon) \int \frac{d^{4}k}{(2\pi)^{4}} \frac{1}{(k^2 + \M^2)^{2+\alpha+\epsilon}}\nn\\
&=& \int \frac{d^4 k}{(2\pi)^{4} } f_{(2+\alpha, -2\alpha)}(\epsilon) \left(\frac{k^2}{\mu^2}+\frac{\M^2}{\mu^2} \right)^{-\epsilon} \frac{1}{(k^2 + \M^2)^{2+\alpha}} \\
J_{-2\alpha}^{\mu \nu} |_\text{den reg}[\theta_{-2\alpha}]  &=& \mu^{2\epsilon} f_{(3+\alpha, -2\alpha)}(\epsilon) \int \frac{d^{4}k}{(2\pi)^{4}} \frac{k^{\mu}k^{\nu}}{(k^2 + \M^2)^{3+\alpha+\epsilon}}    \qquad \nn \\
 &=&\frac14  g^{\mu \nu} \int \frac{d^{4}k}{(2\pi)^4} f_{(3+\alpha, -2\alpha)}(\epsilon) \left(\frac{k^2}{\mu^2}+\frac{\M^2}{\mu^2} \right)^{-\epsilon} \frac{k^2}{(k^2 + \M^2)^{3+\alpha}} \\
 J_{-2\alpha}^{\mu_1 \cdots \mu_4}|_\text{den reg}[\kappa_{-2\alpha}]  (\M^2) 
 &=& \mu^{2\epsilon}  f_{(4+\alpha, -2\alpha)}(\epsilon) \int \frac{d^4k}{(2 \pi)^{4}}  \frac{k^{\mu_1} \cdots k^{\mu_4}}{(k^2 +\M^2)^{4+\alpha}} \nn \\
&=&  \frac{1}{4!}S^{\mu_1 \cdots \mu_4} \int \frac{d^{4} k}{(2 \pi)^{4}}  f_{(4+\alpha, -2\alpha)}(\epsilon) \left(\frac{k^2}{\mu^2}+\frac{\M^2}{\mu^2} \right)^{-\epsilon} \frac{k^4}{(k^2 +\M^2)^{4+\alpha}} \qquad \quad 
\end{eqnarray}
In each case, the function $f_{(n, p)} (\epsilon)$ is introduced for integrals with whose denominator integrand has power $n$, after Feynman parametrisation,  and whose integral has degree of divergence $p$.  These functions should tend smoothly to unity as $\epsilon \to 0$ and are fixed by demanding that the Laurent expansion of the corresponding amplitude in $\epsilon$ converges for all  finite $\epsilon>0$. In other words, the amplitude should only have a simple pole at $\epsilon=0$;  $f_{(n, p)} (\epsilon)$ should cancel any UV poles that go as  $1/\left(\epsilon-p/2\right)$ and any IR poles that emerge for $\epsilon \in \mathbb{N}^+$ even in the massless limit \cite{Horowitz22,Horowitz:2023xyd}.  The mass here refers to the physical masses in the system, as opposed to the mass $\M$ appearing in the ILIs, which also depends on external momenta and Feynman parameters. In practice, the procedure for computing the $f_{(n, p)}(\epsilon)$ requires us to also  compute the integration over the Feynman parameters, as well as the regulated ILIs. In any event, we can read off the form of our $\eta$ regulators as follows
 \begin{equation}
\eta_{-2\alpha}=\frac{f_{(2+\alpha, -2\alpha)}(\epsilon)}{ (y^2+y_0^2)^{\epsilon}} , \qquad \theta_{-2\alpha}=\frac{f_{(3+\alpha, -2\alpha)}(\epsilon)}{ (y^2+y_0^2)^{\epsilon}} , \qquad \kappa_{-2\alpha}=\frac{f_{(4+\alpha, -2\alpha)}(\epsilon)}{ (y^2+y_0^2)^{\epsilon} } \label{denregsol}
\end{equation}
Similar structures can also be derived from our formalism.  To do so, we make the following choice for our seed functions
\be\label{Etadenreg1}
\psi_{-2\alpha}(y, \epsilon)=-\frac{A(\epsilon)}{2(2+\alpha)} \frac{y^{4}}{(y^2+y_0^2)^{2+\alpha+\epsilon}} ,
\qquad
\phi_{-2\alpha}(y,\epsilon)=-\frac{B(\epsilon)}{2(3+\alpha)} \frac{y^{6} }{(y^2+y_0^2)^{3+\alpha+\epsilon}} 
\end{equation}
where, as before, we assume that  the coefficients $A, B$ depend on the regularisation parameter $\epsilon$ and are assumed to behave smoothly at the origin such that $A(0) = B(0)=1$. Plugging this  into \eqref{psi} and \eqref{phi}, we find that
\begin{eqnarray}
\frac{y^2}{y^2+y_0^2} \theta_{-2\alpha}-\frac{2}{2+\alpha}\eta_{-2\alpha} &=& \frac{y^2}{y^2+y_0^2}\left[\frac{2+\alpha+\epsilon}{2+\alpha}\frac{A(\epsilon)}{ (y^2+y_0^2)^\epsilon}\right] -\frac{2}{2+\alpha}\left[\frac{A(\epsilon)}{ (y^2+y_0^2)^\epsilon}\right] \qquad \label{psidenreg}\\
\frac{y^2}{(y^2+y_0^2)} \kappa_{-2\alpha}-\frac{3}{(3+\alpha)}\theta_{-2\alpha} &=&  \frac{y^2}{y^2+y_0^2}\left[\frac{3+\alpha+\epsilon}{3+\alpha}\frac{B(\epsilon)}{ (y^2+y_0^2)^\epsilon}\right] -\frac{3}{3+\alpha}\left[\frac{B(\epsilon)}{ (y^2+y_0^2)^\epsilon}\right]\label{phidenreg}
\end{eqnarray}
This yields \eqref{denregsol} provided we have the following  consistency conditions 
\be
f_{(3+\alpha, -2\alpha)}(\epsilon)=\frac{2+\alpha+\epsilon}{2+\alpha} f_{(2+\alpha, -2\alpha)}(\epsilon), \qquad f_{(4+\alpha, -2\alpha)}(\epsilon)=\frac{3+\alpha+\epsilon}{3+\alpha} f_{(3+\alpha, -2\alpha)}(\epsilon)
\ee
and identify $A(\epsilon)=f_{(2+\alpha, -2\alpha)}(\epsilon)$ and  $B(\epsilon)=f_{(3+\alpha, -2\alpha)}(\epsilon)$. These conditions seem to be violated by some  of the $f_{(n, p)}(\epsilon)$ seen in the literature \cite{Horowitz22}.   This seems to be because Horowitz advocates for a {\it minimal} choice of the $f_{(n, p)}(\epsilon)$, avoiding any additional pre-factors that do not affect the convergence properties of the amplitude. However, it seems that this is too restrictive. It is clear that the amplitude still converges for all finite $\epsilon>0$  if we multiply $f_{(n, p)}(\epsilon)$ by some linear function of $\epsilon$. Contrary to \cite{Bansal:2022juh}, our analysis suggests these non-minimal factors are necessary  if denominator regularisation is to preserve gauge invariance.

\paragraph{Schwinger proper time:}
The next gauge invariant regularisation method  we will consider is Schwinger proper time. In the Schwinger representation of the propagator, the regularised ILIs take the form
\begin{eqnarray}
J_{-2\alpha}|_\text{Schwinger} [\eta_{-2\alpha}]&=& \frac{1}{(1+\alpha)!} \int_0^\infty \frac{d\tau}{\tau} \ \rho(\mu^2 \tau, \epsilon)  \tau^{2 + \alpha} \int \frac{d^4 k}{(2 \pi)^4}  e^{-\tau (k^2+\M^2)}, 
\label{ILIs1ST} \\ 
J_{-2 \alpha}^{ \mu \nu}  |_\text{Schwinger}[\theta_{-2\alpha}]&=& \frac{1}{(2+\alpha)!} \int_0^\infty \frac{d\tau}{\tau} \ \rho(\mu^2 \tau, \epsilon)  \tau^{3 + \alpha} \int \frac{d^4 k}{(2 \pi)^4}  k^\mu k^\nu e^{-\tau (k^2+\M^2)}, 
\label{ILIs2ST} \\ 
J_{-2 \alpha}^{\mu_1\cdots \mu_4} |_\text{Schwinger} [\kappa_{-2\alpha}]&=&  \frac{1}{(3+\alpha)!} \int_0^\infty \frac{d\tau}{\tau} \ \rho(\mu^2 \tau, \epsilon)  \tau^{4 + \alpha} \int \frac{d^4 k}{(2 \pi)^4}  k^{\mu_1} \cdots k^{\mu_4} e^{-\tau (k^2+\M^2)},\quad
 \label{ILIs3ST}
\end{eqnarray}
where  $\rho(u, \epsilon) \to 1$ as $\epsilon \to 0$. For sufficiently large $\epsilon>0$, we also require  $\rho(u, \epsilon) \to 0$  as $u\to 0$, with the rate at which it vanishes being fast enough to render the integrals convergent.    This gives the following form for the regulators
\begin{eqnarray}
\eta_{-2\alpha}&=&  \frac{1}{(1+\alpha)!} \int_0^\infty \frac{d u}{u } \ \rho(u, \epsilon)  \left[u (y^2+y_0^2)\right]^{2 + \alpha}  e^{-u(y^2+y_0^2)}, \label{spt1} \\
\theta_{-2\alpha}&=&  \frac{1}{(2+\alpha)!} \int_0^\infty \frac{d u}{u } \ \rho(u, \epsilon)    \left[u (y^2+y_0^2)\right]^{3 + \alpha}  e^{-u(y^2+y_0^2)}, \label{spt2} \\
\kappa_{-2\alpha}&=&  \frac{1}{(3+\alpha)!} \int_0^\infty \frac{d u}{u } \rho(u, \epsilon) \left[u (y^2+y_0^2)\right]^{4 + \alpha}  e^{-u(y^2+y_0^2)}. \label{spt3}
\end{eqnarray}
To see how this can emerge from our formalism, we first note that our boundary conditions \eqref{eps=0} can be written in the following form
\begin{eqnarray}
\psi_{-2\alpha}(y, 0) &=&- \frac{y^4}{2(2+\alpha)!} \int_0^\infty \frac{d u}{u }   \ u^{2 + \alpha}  e^{-u(y^2+y_0^2)} \\
\phi_{-2\alpha}(y,0) &=& -\frac{y^6}{2(3+\alpha)!}  \int_0^\infty \frac{d u}{u } \  u^{3 + \alpha}  e^{-u(y^2+y_0^2)}  \label{eps=0SPT}
\end{eqnarray}
This inspires the following choice, 
\begin{eqnarray}
\psi_{-2\alpha}(y, \epsilon) &=& -\frac{y^4}{2(2+\alpha)!} \int_0^\infty \frac{d u}{u }  \rho(u, \epsilon)   \ u^{2 + \alpha}  e^{-u(y^2+y_0^2)} \\
\phi_{-2\alpha}(y,\epsilon) &=&- \frac{y^6}{2(3+\alpha)!}  \int_0^\infty \frac{d u}{u } \ \rho(u, \epsilon)   u^{3 + \alpha}  e^{-u(y^2+y_0^2)}  
\end{eqnarray}
After plugging this into \eqref{psi} and \eqref{phi},  we recover the desired regulators \eqref{spt1} to \eqref{spt3}.

\paragraph{A general gauge invariant regulator:}
Finally, we derive a very general gauge invariant regularisation scheme that captures all of the above and more  by making the following choice for our seed functions
\begin{equation}
\psi_{-2\alpha}(y, \epsilon) =\psi_{-2\alpha}(y, 0) \mathcal{F}(\vartheta, \varphi, \epsilon), \qquad \phi_{-2\alpha}(y, \epsilon) =\phi_{-2\alpha}(y, 0) \mathcal{G}(\vartheta, \varphi, \epsilon),
\end{equation}
where $\psi_{-2\alpha}(y, 0) $ and $\phi_{-2\alpha}(y, 0)$ are, of course, given by \eqref{eps=0} and $\vartheta= \ln y^2, ~\varphi= \ln (y^2+y_0^2)$. For sufficiently large $\epsilon >0$, we require that   $y^{-2\alpha}\mathcal{F}(\vartheta, \varphi, \epsilon)$ and $y^{-2\alpha}\mathcal{G}(\vartheta, \varphi, \epsilon)$ should vanish as   $y \to \infty$. We also require that  $\mathcal{F}(\vartheta, \varphi, 0)=\mathcal{G}(\vartheta, \varphi, 0)=1$.  Plugging this into \eqref{psi} and \eqref{phi}, we find that 
\begin{eqnarray}
\frac{y^2}{y^2+y_0^2} \theta_{-2\alpha}-\frac{2}{2+\alpha}\eta_{-2\alpha} &=& \frac{y^2}{y^2+y_0^2}\left[\mathcal{F}- \frac{\del_\varphi \mathcal{F}}{2+\alpha} \right] -\frac{2}{2+\alpha}\left[\mathcal{F}+ \frac{\del_\vartheta \mathcal{F}}{2} \right]\qquad \\
\frac{y^2}{(y^2+y_0^2)} \kappa_{-2\alpha}-\frac{3}{(3+\alpha)}\theta_{-2\alpha} &=&  \frac{y^2}{y^2+y_0^2}\left[\mathcal{G}- \frac{\del_\varphi \mathcal{G}}{3+\alpha} \right] -\frac{3}{3+\alpha}\left[\mathcal{G}+ \frac{\del_\vartheta \mathcal{G}}{3} \right]\qquad 
\end{eqnarray}
For consistency, $\mathcal{F}=\frac13 e^{-3\vartheta+(2+\alpha)\varphi} \del_\vartheta \mathcal{H}$ and $\mathcal{G}=-\frac{1}{2+\alpha}e^{-3\vartheta+(2+\alpha)\varphi}  \del_\varphi \mathcal{H}$, giving the following general form for gauge invariant regulators
\begin{eqnarray}
\eta_{-2\alpha}&=&-\frac16 e^{-3\vartheta+(2+\alpha)\varphi}  \left[\del_\vartheta \mathcal{H}-\del^2_\vartheta \mathcal{H}\right], \\
 \theta_{-2\alpha}&=&-\frac{1}{3(2+\alpha)} e^{-3\vartheta+(2+\alpha)\varphi}  \del_\vartheta \del_\varphi \mathcal{H}, \\
  \kappa_{-2\alpha}&=&-\frac{1}{(2+\alpha)(3+\alpha)} e^{-3\vartheta+(2+\alpha)\varphi} \left[\del_\varphi \mathcal{H}-\del^2_\varphi \mathcal{H}\right]
\end{eqnarray}
Note that $\mathcal{H}(\vartheta, \varphi, 0)=e^{3\vartheta-(2+\alpha)\varphi}$. Meanwhile, for sufficiently large $\epsilon>0$,  we require  $e^{-3 \vartheta+2\varphi}\del_i\mathcal{H} (\vartheta, \varphi, \epsilon) \to 0$ at large $\vartheta\sim \varphi$.

\section{$\eta$ regularisation and anomalies}\label{triangle}

In quantum field theory,  classical symmetries are often broken by quantum effects.  In the case of a global symmetry, this is a physical effect and one that is always expected to occur if the theory is coupled to gravity \cite{Kallosh:1995hi}.   Gauge anomalies, on the other hand, can be fatal, since they allow  new degrees of freedom  to start propagating in the effective field theory, often leading to  violations of  unitarity.  The classic playground for anomalies are chiral models in an even number of spacetime dimensions, in which a gauge field is coupled to massless Dirac fermions \cite{Bertlmann:1996xk}. Classically, there are two conserved currents: a vector current associated with the gauge symmetry and an axial current associated with the chiral symmetry. In two dimensions,  two point functions reveal both symmetries to be anomalous, although somewhat unusually,  this does not lead to any inconsistencies \cite{Jackiw:1984zi}.   In four dimensions, three point functions reveal a chiral anomaly, while the gauge symmetry remains intact  \cite{Adler:1969gk,Bell:1969ts}.  Of course, for non-abelian theories, the go-to regularisation scheme is dimensional regularisation \cite{tHooft:1972tcz} which is tricky to implement for chiral models.   Indeed, the definition of the chiral model makes use of the $\gamma_5$ matrix whose properties are ambiguous when the number of  dimensions is analytically continued  (see, for example,  \cite{Breitenlohner:1977hr,Jegerlehner:2000dz,Tsai:2009it, Tsai:2010aq,Ferrari:2014jqa,Ferrari:2015mha,Ferrari:2016nea, Gnendiger:2017pys,Gnendiger:2017rfh}).

$\eta$ regularisation has the advantage that the dimension of spacetime always remains fixed.  Nevertheless, we have seen how it is capable of capturing the key ingredients of dimensional regularisation and other regularisation schemes that preserve gauge invariance. It is therefore important to better understand how the anomalies of the chiral model should be treated in our panoptic framework. Note that  $\gamma_5$ ambiguities for regularisation schemes in fixed dimension have been considered in \cite{Cynolter:2010ae,Viglioni:2016nqc, Bruque:2018bmy}.   

To simplify our analysis whilst still capturing the important physics, we consider  the chiral Schwinger model.   This is given by the Lagrangian \eqref{lagr} for a single massless Dirac fermion in two spacetime dimensions. Classically, there are two conserved currents: the vector current, $j^V_\mu=\bar \psi \gamma_\mu \psi$ and the axial current,  $j^A_\mu=\bar \psi \gamma_\mu  \gamma_5 \psi$ where we define $\gamma_5=-i \gamma^0 \gamma^1$ in two dimensions.   To investigate the anomalies, we consider correlation functions of these currents, specifically
\be
\Pi^{VV}_{\mu\nu}(p)=\left \langle j^V_\mu(p)j^V_\nu(-p) \right \rangle , \qquad \Pi^{VA}_{\mu\nu}(p)= \left \langle j^V_\mu(p)j^A_\nu(-p) \right \rangle \qquad \Pi^{AA}_{\mu\nu}(p)=\left  \langle j^A_\mu(p)j^A_\nu(-p) \right \rangle 
\ee
The classical Ward identities give $p^\mu \Pi_{\mu\nu}^{\cdots}=p^\nu \Pi_{\mu\nu}^{\cdots }=0$.   These Ward identities are anomalous. It is well known that there is a gauge anomaly coming from the generation of a mass term for the gauge field  \cite{Jackiw:1984zi}.  While this results in the non-conservation of $\Pi^{VV}_{\mu\nu}$ at one loop,  in this particular case  it does not lead to a violation of unitarity.   Here we will  focus on the two point function with mixed vertices, $\Pi^{VA}_{\mu\nu}$ for which we expect only the axial part to be anomalous
$p^\mu \Pi^{VA}_{\mu\nu}=0, \ p^\nu \Pi^{AV}_{\mu\nu}  \neq 0$. At one loop, we have that 
\be
i\Pi^{VA }_{\mu\nu}=- \int \frac{d^2 k}{(2\pi)^2} \Tr  \left(  \gamma_\mu  \frac{i}{\slashed{k}-\slashed{p}+i \epsilon } \gamma_\nu \gamma_5 \frac{i}{\slashed{k}+i \epsilon }\right)
\ee
The first step in $\eta$ regularisation is to cast this into an ILI form. To this end we perform a Wick rotation $k_0 \to ik_2$,  $\gamma^0 \to -i \gamma^2$ and $\gamma_5 \to \gamma^1 \gamma^2$  to Euclidean signature along with a  Feynman parametrisation, giving 
\be
\Pi^{VA}_{\mu\nu}=\int \frac{d^2 k}{(2\pi)^2} \int_0^1 dx  \Tr  \left(\gamma_5 \gamma_\sigma \gamma_\mu \gamma_\rho \gamma_\nu\right)  \frac{(k-xp+xp)^\sigma (k-xp+(x-1)p)^\rho }{\left[(k-xp)^2+\M^2\right]^2}
\ee
where $\M^2=x(1-x)p^2$. At this stage, the temptation might be to pull the trace operator $  \Tr  \left(\gamma_5 \gamma_\sigma \gamma_\mu \gamma_\rho \gamma_\nu\right) $ out of the integral and write the correlation function as a product of the trace operator and the corresponding ILIs. However,  one cannot pull the trace through the divergent integral without leaving it unchanged  \cite{Bruque:2018bmy}.  The correct thing to do is to leave trace operator {\it inside} the integral, decompose it in terms of the metric and express the final result in terms of the ILIs.  To this end, we note that
\be
\Tr  \left(\gamma_5 \gamma_\sigma \gamma_\mu \gamma_\rho \gamma_\nu\right)=2(-\epsilon_{\sigma\mu} g_{\rho\nu}+\epsilon_{\sigma\rho} g_{\mu\nu}-\epsilon_{\sigma\nu} g_{\mu \rho}-\epsilon_{\mu\rho} g_{\sigma\nu}+\epsilon_{\mu\nu} g_{\sigma\rho}-\epsilon_{\rho\nu} g_{\sigma\mu})
\ee
where $\epsilon_{\mu\nu}$ is the totally antisymmetric Levi-Civita tensor in two dimensions. Using the rotational symmetry of the integral, the correlation function can be written as 
\be
\Pi^{VA}_{\mu\nu}=-4 \int \frac{d^2 k}{(2\pi)^2} \int_0^1 dx \  \epsilon_{\sigma\nu }\left[ \mathcal{J}_{0}^\sigma{} _\mu-\frac12 \mathcal{J}_{0}^\lambda{}_\lambda \delta^\sigma _\mu  -x(1-x)  \left(p^\sigma p_\mu-\frac12 p^2 \delta^\sigma _\mu \right)  \mathcal{J}_{-2}\right]
\ee
where 
\be 
 \mathcal{J}_{-2\alpha }^{\mu\nu}=\frac{(k-xp)^\mu (k-xp)^\nu}{\left[(k-xp)^2+\M^2\right]^{2+\alpha}}, \qquad \mathcal{J}_{-2\alpha}=\frac{1}{\left[(k-xp)^2+\M^2\right]^{1+\alpha}}
\ee
To put this in ILI form, we use the fact that $ \mathcal{J}_{0}^\lambda{}_\lambda =  \mathcal{J}_{0}-x(1-x)p^2 \mathcal{J}_{-2} $ so that we can write
\be \label{piVA}
\Pi^{VA}_{\mu\nu}=-4\int_0^1 dx \  \epsilon_{\sigma\nu }\left[ {J}_{0}^\sigma{} _\mu[\theta_0](\M^2)-\frac12{J}_{0}[\eta_0] (\M^2)\delta^\sigma _\mu -x(1-x)  \left(p^\sigma p_\mu-p^2 \delta^\sigma _\mu \right) {J}_{-2} \right]
\ee
Here we have introduced the ILIs in two dimensions along with their corresponding regulators
\be
{J}^{\mu_1 \cdots \mu_2}_{-2\alpha}[\eta](\M^2) = \int \frac{d^2 k}{(2\pi)^2}\frac{k^{\mu_1} \cdots k^{\mu_s}}{\left[k^2+\M^2\right]^{1+\frac{s}{2}+\alpha}} \eta\left(\frac{k}{\mu}, \epsilon\right)
\ee
Note that the final ILI in  \eqref{piVA} is convergent with $ {J}_{-2}[1]=1/4 \pi \M^2$. We now use momentum routing in two dimensions to derive a consistency condition for the logarithmically divergent ILIs, 
\begin{equation}
 \int \frac{d^2 k}{(2 \pi)^2} \frac{\partial}{\partial k^{\mu}  }        \left[  \frac{k_{\nu}}{k^2 +\M^2}\rho\left(\frac{k}{\mu}, \epsilon\right)\right]=0 \implies {J}_{0}^{\mu\nu}[\theta_0] (\M^2)=\frac12{J}_{0}[\eta_0] (\M^2) g^{\mu\nu}
\end{equation}
Plugging all of this into  \eqref{piVA}  we arrive at the final expression, 
\be
\Pi^{VA}_{\mu\nu}=\frac{\epsilon_{\sigma\nu }}{\pi}  \left(\hat p^\sigma \hat p_\mu- \delta^\sigma _\mu \right)
\ee
where $\hat p^\mu=p^\mu/p$.  It is easy to check that this correlation function generates an axial anomaly but not a vector anomaly, consistent with our expectations,
\be
p^\mu\Pi^{VA}_{\mu\nu}=0, \qquad p^\nu\Pi^{VA}_{\mu\nu}=-\frac{\epsilon_{\mu\nu} p^\nu}{\pi}
\ee
It is now clear  $\eta$ regularisation avoids some of the confusion and ambiguity associated with analytically continued dimensions in chiral theories.  However, it is important that $\eta$ regularisation is properly implemented. To recover the correct result for the anomaly, it was crucial for us to fully decompose the  integrand in the correlation function into the integrands of the ILIs, taking all the relevant contractions with the metric  {\it inside} the integral. Had we pulled the trace operator out of the integral before doing these decompositions and contractions, we would have got the wrong result. This is because one cannot move metric contractions through a divergent integral without generating possible corrections  \cite{Bruque:2018bmy}.

\section{Discussion} \label{discussion}
Originally inspired by regularisation techniques in number theory, $\eta$ regularisation has provided us with a tool to investigate regularisation schemes in quantum field theory in a systematic way, at least at one loop.  Here we have shown how it can be used to ``solve" gauge consistency conditions,  delivering old and new gauge invariant regularisation schemes.  This includes dimensional regularisation, denominator regularisation, Schwinger proper time and a generalisation that captures all of the above and more.  We have also shown how the careful implementation of $\eta$ regularisation yields the correct results for anomalies in chiral theories. 

The extension of $\eta$ regularisation to higher loops is of paramount importance.  Given the role of ILIs in developing the formalism, we expect an extension to higher loops to closely follow the analysis of Wu \cite{Wu03,Wu04,Wu14}. These extensions are non-trivial, not least because of the presence of overlapping divergences that begin at two loops.  Extending $\eta$ regularisation to higher loops would also allow us to investigate unitarity, potentially deriving and solving new consistency conditions associated with same.  We would also like to use $\eta$ regularisation to ``solve" for regularisation schemes that preserve supersymmetry. 

However, our ultimate goal is to use $\eta$ regularisation to gain a window into how nature protects us from infinity.  For example,  superstring theory is said to be UV finite \cite{Atick:1987zt,Mandelstam:1991tw,Berkovits:2004px,Sen:2015cxs}.   When we carefully take the field theory limit, this UV finiteness should manifest itself in some way. This question was explored in \cite{Tourkine:2013rda} where the heavy string oscillator modes provide the appropriate counter terms in the field theory limit.   Here we imagine another perspective, whereby the field theory inherits some sort of stringy regulator.  What would this regulator look like and can it be generalised without spoiling its most important ingredients?  Presumably, if string theory is consistent, a string inspired regulator should satisfy all the necessary consistency conditions anticipated from bottom up investigations like the one presented here.  Furthermore, it is well known that worldsheet modular invariance is key to ensuring that string theory remains UV finite.  This suggests that  string inspired regulators might be expected to feature this in some way, perhaps through some subtle UV/IR mixing that leads to unexpected cancellations as in \cite{Abel:2021tyt,Abel:2023hkk,Abel:2024twz}. 
\begin{appendices}
\section{Undoing the Feynman parametrisation with $\eta$ factors}
In the main text, we introduced our $\eta$ regulators after performing a range of manipulations on the integrals, most notably, Feynman parametrisations.  Such manipulations are not valid for divergent integrals so one is implicitly assuming that some form of regularisation is always present and that this will result in the proposed $\eta$ regularisation of the ILIs at the end of the calculation.  We shall now investigate how the regularisation scheme might look prior to Feynman parametrisation for certain integrals.

To this end, we consider a contribution to a correlation function coming from a regularised scalar ILI
\be
\mathcal{A}= (n-1)!   \int_x \int_k \frac{1}{(k^2+\M^2)^n} \eta\left(\frac{k}{\mu}, \epsilon\right) 
\ee
Here $\int_x= \int_0^1 d x_1 \cdots \int_0^1 dx_n \delta(\sum_i x_i-1)$ denotes the integration over $n$ Feynman parameters, $x_i$, subject to the constraint $\sum_i x_i=1$.  Furthermore,  $\M^2=\sum_i x_i(p_i^2+m_i^2)-\left(\sum_i x_i p_i \right)^2$ is a function of these parameters, the particle masses, $m_i$ and external momenta, $p_i$.  $\int_k$ denotes the integration over the loop momentum, which could be in any number of dimensions.  

To undo the Feynman parametrisation, it is convenient to rewrite the integral in Schwinger form. To do so, we first use momentum routing to shift the loop momentum 
\begin{eqnarray}
\mathcal{A} &=& (n-1)!\int_x  \int_k\frac{1}{((k+\sum_i x_i p_i)^2+M^2)^n} \eta\left(\frac{|k+\sum_i x_i p_i|}{\mu}, \epsilon\right) \\
&=& (n-1)! \int_x  \int_k\frac{1}{(\sum_i x_i \Delta_i )^n} \eta\left(\frac{|k+\sum_i x_i p_i|}{\mu}, \epsilon\right)
\end{eqnarray}
where $\Delta_i=(k+p_i)^2+m_i^2$,  before integrating in a Schwinger parameter, $s$
\be
\mathcal{A} = \int_0^\infty ds \ s^{n-1} \int_x \int_k e^{-s(\sum_i x_i \Delta_i )} \eta\left(\frac{|k+\sum_i x_i p_i|}{\mu}, \epsilon\right)
\ee
We now introduce a set of $n$ variables $s_i=s x_i$ allowing us to combine the integration of the Schwinger parameter and the Feynman parameters, 
\be
\mathcal{A} = \int_0^\infty  ds_1 \cdots \int_0^\infty  ds_n  \int_k e^{-(\sum_i s_i \Delta_i )} \eta\left(\frac{|k+\frac{\sum_i s_i p_i}{\sum_i s_i}|}{\mu}, \epsilon\right)
\ee
Finally, we absorb the $\Delta_i$ into a redefinition of the $s_i \to s_i/\Delta_i$ so that we have
\be
\mathcal{A} =\int_k \frac{1}{\Delta_1 \cdots \Delta_n}\tilde \eta
\ee
where 
\be
\tilde \eta = \int_0^\infty  ds_1 e^{-s_1} \cdots \int_0^\infty  ds_n  e^{- s_n } \eta\left(\frac{\left|k+\frac{\sum_i s_i p_i/\Delta_i}{\sum_i s_i/\Delta_i}\right|}{\mu}, \epsilon\right)
\ee
This is the form of regularisation factor prior to Feynman parametrisation. Note that $\tilde \eta \to 1$ and $\eta \to 1$, as of course it should. 

\end{appendices}

\section*{Acknowledgements}
We are grateful to Paolo Di Vecchia, Pete Millington, Killian Moehling, Benjamin Muntz and Paul Saffin for useful comments and discussions. RGCS was supported by a Bell Burnell Studentship and AP  by STFC consolidated grant number ST/T000732/1.  For the purpose of open access, the authors have applied a CC BY public copyright licence to any Author Accepted Manuscript version arising. No new data were created during this study.

\bibliography{QFTpapers}

\providecommand{\href}[2]{#2}\begingroup\raggedright\begin{thebibliography}{10}

\bibitem{Bertlmann:1996xk}
R.~A. Bertlmann, \emph{{Anomalies in quantum field theory}}.
\newblock 1996.

\bibitem{Schwartz14}
M.~D. Schwartz, \emph{{Quantum Field Theory and the Standard Model}}.
\newblock Cambridge University Press, 3, 2014.

\bibitem{Weinberg95a}
S.~Weinberg, \emph{{The Quantum theory of fields. Vol. 1: Foundations}}.
\newblock Cambridge University Press, 6, 2005,
  \href{http://dx.doi.org/10.1017/CBO9781139644167}{10.1017/CBO9781139644167}.

\bibitem{Fujikawa:2004cx}
K.~Fujikawa and H.~Suzuki, \emph{{Path integrals and quantum anomalies}}.
\newblock 2004,
  \href{http://dx.doi.org/10.1093/acprof:oso/9780198529132.001.0001}{10.1093/acprof:oso/9780198529132.001.0001}.

\bibitem{bilal2008lectures}
A.~Bilal, \emph{Lectures on anomalies},  2008.

\bibitem{Adler:1969gk}
S.~L. Adler, \emph{{Axial vector vertex in spinor electrodynamics}},
  \href{http://dx.doi.org/10.1103/PhysRev.177.2426}{\emph{Phys. Rev.} {\bf 177}
  (1969) 2426--2438}.

\bibitem{Bell:1969ts}
J.~S. Bell and R.~Jackiw, \emph{{A PCAC puzzle: $\pi^0 \to \gamma \gamma$ in
  the $\sigma$ model}}, \href{http://dx.doi.org/10.1007/BF02823296}{\emph{Nuovo
  Cim. A} {\bf 60} (1969) 47--61}.

\bibitem{tHooft:1972tcz}
G.~'t~Hooft and M.~J.~G. Veltman, \emph{{Regularization and Renormalization of
  Gauge Fields}},
  \href{http://dx.doi.org/10.1016/0550-3213(72)90279-9}{\emph{Nucl. Phys. B}
  {\bf 44} (1972) 189--213}.

\bibitem{PadillaSmith24}
A.~Padilla and R.~G.~C. Smith, \emph{{Smoothed asymptotics: from number theory
  to QFT}},  \href{http://arxiv.org/abs/2401.10981}{{\tt 2401.10981}}.

\bibitem{Tao11}
T.~Tao, \emph{Compactness and contradiction}.
\newblock American Mathematical Soc., 2013.

\bibitem{Wu03}
Y.-L. Wu, \emph{{Symmetry principle preserving and infinity free regularization
  and renormalization of quantum field theories and the mass gap}},
  \href{http://dx.doi.org/10.1142/S0217751X03015222}{\emph{Int. J. Mod. Phys.
  A} {\bf 18} (2003) 5363--5420},
  [\href{http://arxiv.org/abs/hep-th/0209021}{{\tt hep-th/0209021}}].

\bibitem{Wu04}
Y.-L. Wu, \emph{{Symmetry preserving loop regularization and renormalization of
  QFTs}}, \href{http://dx.doi.org/10.1142/S0217732304015361}{\emph{Mod. Phys.
  Lett. A} {\bf 19} (2004) 2191--2204},
  [\href{http://arxiv.org/abs/hep-th/0311082}{{\tt hep-th/0311082}}].

\bibitem{Wu14}
Y.-L. Wu, \emph{{Quantum Structure of Field Theory and Standard Model Based on
  Infinity-free Loop Regularization/Renormalization}},
  \href{http://dx.doi.org/10.1142/S0217751X14300075}{\emph{Int. J. Mod. Phys.
  A} {\bf 29} (2014) 1430007}, [\href{http://arxiv.org/abs/1312.1403}{{\tt
  1312.1403}}].

\bibitem{Battistel:1998sz}
O.~A. Battistel, A.~L. Mota and M.~C. Nemes, \emph{{Consistency conditions for
  4-D regularizations}},
  \href{http://dx.doi.org/10.1142/S0217732398001686}{\emph{Mod. Phys. Lett. A}
  {\bf 13} (1998) 1597--1610}.

\bibitem{Wu05}
Y.-L. Ma and Y.-L. Wu, \emph{{Anomaly and anomaly-free treatment of QFTs based
  on symmetry-preserving loop regularization}},
  \href{http://dx.doi.org/10.1142/S0217751X0603309X}{\emph{Int. J. Mod. Phys.
  A} {\bf 21} (2006) 6383--6456},
  [\href{http://arxiv.org/abs/hep-ph/0509083}{{\tt hep-ph/0509083}}].

\bibitem{Horowitz:2022rpp}
W.~A. Horowitz and J.~F.~D. Plessis, \emph{{Finite system size correction to
  NLO scattering in \ensuremath{\phi}4 theory}},
  \href{http://dx.doi.org/10.1103/PhysRevD.105.L091901}{\emph{Phys. Rev. D}
  {\bf 105} (2022) L091901}, [\href{http://arxiv.org/abs/2203.01259}{{\tt
  2203.01259}}].

\bibitem{Horowitz22}
W.~A. Horowitz, \emph{{Denominator Regularization in Quantum Field Theory}},
  \href{http://arxiv.org/abs/2209.02820}{{\tt 2209.02820}}.

\bibitem{Bansal:2022juh}
A.~Bansal, N.~Mahajan and D.~Mishra, \emph{{More on Denominator Regularization
  in Quantum Field Theory}},  \href{http://arxiv.org/abs/2211.12284}{{\tt
  2211.12284}}.

\bibitem{Horowitz:2023xyd}
W.~A. Horowitz, \emph{{Jets in e+A SIDIS and Denominator Regularization}},
  \href{http://dx.doi.org/10.1088/1742-6596/2586/1/012019}{\emph{J. Phys. Conf.
  Ser.} {\bf 2586} (2023) 012019}.

\bibitem{Kallosh:1995hi}
R.~Kallosh, A.~D. Linde, D.~A. Linde and L.~Susskind, \emph{{Gravity and global
  symmetries}}, \href{http://dx.doi.org/10.1103/PhysRevD.52.912}{\emph{Phys.
  Rev. D} {\bf 52} (1995) 912--935},
  [\href{http://arxiv.org/abs/hep-th/9502069}{{\tt hep-th/9502069}}].

\bibitem{Jackiw:1984zi}
R.~Jackiw and R.~Rajaraman, \emph{{Vector Meson Mass Generation Through Chiral
  Anomalies}}, \href{http://dx.doi.org/10.1103/PhysRevLett.54.1219}{\emph{Phys.
  Rev. Lett.} {\bf 54} (1985) 1219}.

\bibitem{Breitenlohner:1977hr}
P.~Breitenlohner and D.~Maison, \emph{{Dimensional Renormalization and the
  Action Principle}}, \href{http://dx.doi.org/10.1007/BF01609069}{\emph{Commun.
  Math. Phys.} {\bf 52} (1977) 11--38}.

\bibitem{Jegerlehner:2000dz}
F.~Jegerlehner, \emph{{Facts of life with gamma(5)}},
  \href{http://dx.doi.org/10.1007/s100520100573}{\emph{Eur. Phys. J. C} {\bf
  18} (2001) 673--679}, [\href{http://arxiv.org/abs/hep-th/0005255}{{\tt
  hep-th/0005255}}].

\bibitem{Tsai:2009it}
E.-C. Tsai, \emph{{Gauge Invariant Treatment of $\gamma_{5}$ in the Scheme of
  't Hooft and Veltman}},
  \href{http://dx.doi.org/10.1103/PhysRevD.83.025020}{\emph{Phys. Rev. D} {\bf
  83} (2011) 025020}, [\href{http://arxiv.org/abs/0905.1550}{{\tt 0905.1550}}].

\bibitem{Tsai:2010aq}
E.-C. Tsai, \emph{{Maintaining Gauge Symmetry in Renormalizing Chiral Gauge
  Theories}}, \href{http://dx.doi.org/10.1103/PhysRevD.83.065011}{\emph{Phys.
  Rev. D} {\bf 83} (2011) 065011}, [\href{http://arxiv.org/abs/1012.3501}{{\tt
  1012.3501}}].

\bibitem{Ferrari:2014jqa}
R.~Ferrari, \emph{{Managing $\gamma_5$ in Dimensional Regularization and ABJ
  Anomaly}},  \href{http://arxiv.org/abs/1403.4212}{{\tt 1403.4212}}.

\bibitem{Ferrari:2015mha}
R.~Ferrari, \emph{{Managing $\gamma_5$ in Dimensional Regularization II: the
  Trace with more $\gamma_{5}\prime s$}},
  \href{http://dx.doi.org/10.1007/s10773-016-3211-8}{\emph{Int. J. Theor.
  Phys.} {\bf 56} (2017) 691--705},
  [\href{http://arxiv.org/abs/1503.07410}{{\tt 1503.07410}}].

\bibitem{Ferrari:2016nea}
R.~Ferrari, \emph{{$\gamma_5$ in Dimensional Regularization: a Novel
  Approach}},  \href{http://arxiv.org/abs/1605.06929}{{\tt 1605.06929}}.

\bibitem{Gnendiger:2017pys}
C.~Gnendiger et~al., \emph{{To ${d}$, or not to ${d}$: recent developments and
  comparisons of regularization schemes}},
  \href{http://dx.doi.org/10.1140/epjc/s10052-017-5023-2}{\emph{Eur. Phys. J.
  C} {\bf 77} (2017) 471}, [\href{http://arxiv.org/abs/1705.01827}{{\tt
  1705.01827}}].

\bibitem{Gnendiger:2017rfh}
C.~Gnendiger and A.~Signer, \emph{{$\gamma_{5}$ in the four-dimensional
  helicity scheme}},
  \href{http://dx.doi.org/10.1103/PhysRevD.97.096006}{\emph{Phys. Rev. D} {\bf
  97} (2018) 096006}, [\href{http://arxiv.org/abs/1710.09231}{{\tt
  1710.09231}}].

\bibitem{Cynolter:2010ae}
G.~Cynolter and E.~Lendvai, \emph{{Note on triangle anomaly with improved
  momentum cutoff}},
  \href{http://dx.doi.org/10.1142/S021773231103595X}{\emph{Mod. Phys. Lett. A}
  {\bf 26} (2011) 1537--1545}, [\href{http://arxiv.org/abs/1012.4648}{{\tt
  1012.4648}}].

\bibitem{Viglioni:2016nqc}
A.~C.~D. Viglioni, A.~L. Cherchiglia, A.~R. Vieira, B.~Hiller and M.~Sampaio,
  \emph{{$\gamma_{5}$ algebra ambiguities in Feynman amplitudes: Momentum
  routing invariance and anomalies in $D=4$ and $D=2$}},
  \href{http://dx.doi.org/10.1103/PhysRevD.94.065023}{\emph{Phys. Rev. D} {\bf
  94} (2016) 065023}, [\href{http://arxiv.org/abs/1606.01772}{{\tt
  1606.01772}}].

\bibitem{Bruque:2018bmy}
A.~M. Bruque, A.~L. Cherchiglia and M.~P\'erez-Victoria, \emph{{Dimensional
  regularization vs methods in fixed dimension with and without $\gamma_5$}},
  \href{http://dx.doi.org/10.1007/JHEP08(2018)109}{\emph{JHEP} {\bf 08} (2018)
  109}, [\href{http://arxiv.org/abs/1803.09764}{{\tt 1803.09764}}].

\bibitem{Atick:1987zt}
J.~J. Atick, G.~W. Moore and A.~Sen, \emph{{CATOPTRIC TADPOLES}},
  \href{http://dx.doi.org/10.1016/0550-3213(88)90322-7}{\emph{Nucl. Phys. B}
  {\bf 307} (1988) 221--273}.

\bibitem{Mandelstam:1991tw}
S.~Mandelstam, \emph{{The n loop string amplitude: Explicit formulas,
  finiteness and absence of ambiguities}},
  \href{http://dx.doi.org/10.1016/0370-2693(92)90961-3}{\emph{Phys. Lett. B}
  {\bf 277} (1992) 82--88}.

\bibitem{Berkovits:2004px}
N.~Berkovits, \emph{{Multiloop amplitudes and vanishing theorems using the pure
  spinor formalism for the superstring}},
  \href{http://dx.doi.org/10.1088/1126-6708/2004/09/047}{\emph{JHEP} {\bf 09}
  (2004) 047}, [\href{http://arxiv.org/abs/hep-th/0406055}{{\tt
  hep-th/0406055}}].

\bibitem{Sen:2015cxs}
A.~Sen, \emph{{Ultraviolet and Infrared Divergences in Superstring Theory}},
  \href{http://arxiv.org/abs/1512.00026}{{\tt 1512.00026}}.

\bibitem{Tourkine:2013rda}
P.~Tourkine, \emph{{Tropical Amplitudes}},
  \href{http://dx.doi.org/10.1007/s00023-017-0560-7}{\emph{Annales Henri
  Poincare} {\bf 18} (2017) 2199--2249},
  [\href{http://arxiv.org/abs/1309.3551}{{\tt 1309.3551}}].

\bibitem{Abel:2021tyt}
S.~Abel and K.~R. Dienes, \emph{{Calculating the Higgs mass in string theory}},
  \href{http://dx.doi.org/10.1103/PhysRevD.104.126032}{\emph{Phys. Rev. D} {\bf
  104} (2021) 126032}, [\href{http://arxiv.org/abs/2106.04622}{{\tt
  2106.04622}}].

\bibitem{Abel:2023hkk}
S.~Abel, K.~R. Dienes and L.~A. Nutricati, \emph{{Running of gauge couplings in
  string theory}},
  \href{http://dx.doi.org/10.1103/PhysRevD.107.126019}{\emph{Phys. Rev. D} {\bf
  107} (2023) 126019}, [\href{http://arxiv.org/abs/2303.08534}{{\tt
  2303.08534}}].

\bibitem{Abel:2024twz}
S.~Abel, K.~R. Dienes and L.~A. Nutricati, \emph{{A New Non-Renormalization
  Theorem from UV/IR Mixing}},  \href{http://arxiv.org/abs/2407.11160}{{\tt
  2407.11160}}.

\end{thebibliography}\endgroup
\end{document}